\documentstyle[12pt]{article}
\begin{document}

\begin{titlepage}
\begin{flushright}
King's College KCL-TH-96-18\\
hepth@xxx/9611074\\
\today
\end{flushright}

\bigskip\bigskip\begin{center} {\bf
\Large{Is N=4 Yang-Mills Theory Soluble?}\footnote{Talk given by PCW at The
Quantum Gravity Seminar June 12-19, 1995, Moscow and at the Imperial College
Workshop on "Gauge Theories, Applied Supersymmetry and Quantum Gravity", 5-10
July 1996}}
\end{center} \vskip 1.0truecm

\centerline{\bf P.S. Howe}
\vskip 5mm
\centerline{and}
\vskip 5mm
\centerline{\bf P.C. West}
\vskip 5mm
\centerline{Department of Mathematics}
\centerline{King's College, London}
\vskip5mm

\bigskip \nopagebreak \begin{abstract}
\noindent

\end{abstract}
The superconformal properties of $N=4$ Yang-Mills theory are most naturally
studied using the formalism of harmonic superspace. Superconformal
invariance is shown to imply that the Green's functions of analytic operators
are invariant holomorphic sections of a line bundle on a product of certain
harmonic superspaces and it is  argued
that the theory is soluble for a class of such
operators.

\end{titlepage}

Theories with extended rigid supersymmetry are the most
symmetric consistent quantum
field theories that we know which do not include gravity.
Since field theories that involve
spin $3 \over 2$ particles require spin 2  particles for classical
consistency,  we must,
in the absence of
gravity, restrict our considerations to spins
one and
less.  The celebrated no-go theorem ensures that if we want to
combine internal and Poincar\'e group  symmetries in a non-trivial way then
we must include supersymmetry.  Supersymmetric theories are
classified by the number, $N$, of supercharges they have.  In four dimensions
theories
with more than four supercharges contain spins greater than one and so
must, for consistency, be supergravity theories. As such,
the maximally supersymmetric  theory with only rigid supersymmetry can
have at most four supersymmetries. Such a theory has a unique particle
content given by 1
spin one, 4 spin $1\over2$ and 6 spin 0 all in the adjoint representation of a
gauge group.
This theory is called the
$N = 4$ Yang-Mills theory and its action is uniquely determined once
the coupling constant and gauge group are specified.  Given that this
theory is uniquely picked  out on grounds of symmetry, one may hope
that its Green's functions can also be determined by its
symmetries. Long ago, it was shown that the $N=4$ Yang-Mills theory
\cite{finite} and
a large class of $N=2$ rigid supersymmetric theories \cite {hsw} are
conformally invariant even as quantum theories. Hence the $N=4$
Yang-Mills theory actually has $N=4$ superconformal symmetry.
\par
Although a systematic understanding of how
to compute non-perturbative effects in quantum field theory is still
lacking, in the 1980's spectacular progress \cite {bpz} was made when
it was shown that one could solve  a
large class of two-dimensional conformal models, the so-called
minimal models. Solve in this context means that one can  determine
explicitly the
Green's functions of these theories.
\par
The most likely four-dimensional theory for which
one could hope to achieve a similar
success is the $N=4$ Yang-Mills theory as it is
 conformally invariant
and uniquely picked out on grounds of symmetry.   It is well-known
however, that for a generic conformal theory, conformal invariance
determines only the two- and three-point Green's functions.  Given two
possible Green's functions for a given set of operators, their ratio
will be a conformal invariant.  Hence, the existence of conformal
invariants will, in the absence
of other constraints, imply the theory is not soluble.
It is well-known that
for four and more space-time points conformal invariants
do exist. One may hope that supersymmetry by itself constrains the
number of invariants.  However, one finds that
the basic building block of conformal invariants,  the four point
cross-ratio, generalises in a straightforward way to Minkowski superspace.
Hence,
at  first sight, it would not seem likely that one could use the
superconformal invariance of $N=4$ Yang-Mills theory to  solve for its
Green's functions explicitly.

Almost all supersymmetric theories of interest admit a
superfield description based on
superfields that are constrained.  For example, the Wess-Zumino
multiplet and the superspace field strengths
of the $N = 1$ and 2 Yang-Mills theories are described by chiral
superfields. One encouraging sign for $N=4$ is that,  in superconformal
theories, one can determine \cite {con} the
anomalous dimension of any chiral operator if one knows its weight
under
$R$-transformations and  one can also determine the
Green's functions which depend only on superfields of a given
chirality \cite {world}. This is not the case  for operators and
correlators in   generic conformal field theories.

A chiral superfield, $\varphi$, can be defined on Minkowski
superspace which has cooordinates
\begin{equation}
 x^{\alpha \dot\alpha}, \theta^{ \alpha i}, \bar\theta^{\dot \alpha}_i; \ \
 \alpha, \dot\alpha
=1, 2 \ \ , i = 1, \dots , N.
\end{equation}
and is subject to the constraint $\bar D_{\dot \alpha}^i \varphi = 0$
where $\bar D_{\dot \alpha}^i$ is the
superspace dotted spinor derivative.  The meaning
of this constraint becomes
clear if we make a coordinate change to
$(y^\mu = x^\mu - i \theta^{\alpha i} (\sigma^\mu)_{\alpha
\dot \alpha} \bar\theta ^{\dot \alpha}_i \ \ , \theta^{\alpha i} \ \ ,
 \bar\theta^{\dot \alpha}_ i)$ whereupon $ \bar D_{\dot \alpha}^i \varphi=0$
 becomes simply
${\partial\over \partial\bar\theta^{\dot \alpha}_i} \varphi = 0$ so that
$\varphi$ depends only on $y^\mu$ and $\theta^{\alpha i}$, the
coordinates of chiral superspace.  The ability to solve a given chiral
sector of the theory is a consequence of the fact that there exist no
superconformal invariants which are Grassmann even, non-nilpotent,
and which depend only on  $y^\mu_a,
\theta^{\alpha i}_a$ where $a$ labels different points of chiral superspace
corresponding to the different operators in a Green's function
\cite {world}.
\par
The above discussion can be given a more mathematically sophisticated
formulation.   The complex conformal group is $SL(4, {\bf C}),$
while the complex superconformal group is $SL(4 \vert N, {\bf
C}).$  We can regard complex space-time and complex  Minkowski
superspace as coset spaces of these groups divided by  suitable
subgroups,
$H$ and $H_s$ respectively.  Chiral superspace is the coset space
$H_{sc}\backslash SL(4\vert N, {\bf C}) $ where  $H_s$ is a subgroup of
$H_{sc}$.

 The fields of interest to us transform under induced representations.   We
recall that given a coset space $H\backslash G$, a field transforming
under an induced respresentation is a field defined on the group which
satisfies the condition $\phi (hg)= D(h)\phi(g) \forall \ \  g\in G, \ \ h
\in H $ and $D (h)$ is the matrix respresentation
carried by
$\phi.$  The action of the group is then defined to be $U(g_1) \phi (g)
= \phi^\prime (g) =  \phi (g g_1).$  Although we can work with
fields defined on the group, it is often more convenient to choose
coset representatives $s(x)$ and consider
$\phi$ to depend only on x by writing every $g$ in the form
 $g = h s(x)$.
The action of the group then becomes
\begin{equation}
U (g_1) \phi (s(x))= D(h_1) \phi (s(x^\prime)).
\end{equation}
where $h_1$ is an appropriate element of the isoptropy group depending on $g_1$
and $s(x)$ which is given by $s(x)g_1=h_1s(x^\prime)$.
Carrying out this proceedure for a chiral
superspace we find that $\varphi$ depends on  only the coordinates
$y^\mu, \theta^{\alpha i}.$  We can recover the more usual
formulation of chiral superfields defined on Minkowski superspace by
using only
$h\in H_s$ instead of
$H_{sc}$ to gauge away parts of any $g \in G.$ Then $\varphi$
depends on
$y^\mu, \theta^{\alpha i}, \theta^{\dot \alpha}_i$, but is subject to
$\varphi(h g) = D (h) \varphi (g)=0$ for $h \in H_{sc}$ but $h
\notin H_s.$  This latter  constraint is none other than the chiral
constraint $\bar D_{\dot \alpha}^i\varphi=0$. Hence, using the coset
representatives for chiral superspace automatically solves the chiral
constraint.

The $N = 4$ Yang-Mills theory has a superspace  formulation \cite
{hart} based on harmonic superspace \cite {harm}, the latter being an
extension of Minkowski superspace to include a Grassmann even internal
space which is the coset space of two Lie groups. It is usual to take
the superfields to be defined initially on the product of Minkowski superspace
and
the larger of the internal groups associated with the internal coset
and subject them  to
constraints.  For our purposes, it will be more useful to follow the
discussion of the chiral superfield given above where we in effect
solved these constraints.  The complex superconformal group for
$N$-extended supersymmetry is $SL(4 \vert N; {\bf C})$  and, if we
divide by an appropriate subgroup $H_N$, we can construct the harmonic
superspace analogue of chiral superspace. This superspace is not harmonic
superspace itself, but a new superspace which is usually called analytic
superspace. We choose coset representatives
which can, by appropriate swapping of rows and columns, be
parameterised by the elements of
$GL({N\over 2}\vert 2, {\bf C})$
 matrix, $X$ \cite {world},
\begin{equation}
X = \ \left(\matrix { x \ \ &\lambda\cr
\pi \ \ &y\cr}\right)
\end{equation}
The matrix $x$ contains the coordinates  of space-time, $\lambda$ and
$\pi$ are spinors and $y$ are the  coordinates of the internal
coset space.  The transformation \cite {world}\cite {hw2}of $X$ under
an  arbitary group element
$g_1\in \ \ S L (4\vert N ; {\bf C})$, which can be written $g_1
= \left(\matrix {A \ \ \ &B\cr
C \ \ \ &D\cr}\right)$
where $A, B, C, D \  \in GL({N\over 2}\vert 2, {\bf C})$
is
\begin{equation}
 X \rightarrow X^\prime \ = \ \ \ [A + C X]^{- 1} \ \ [B + X D]
 \end{equation}
The superfields of interest to us transform under induced
representations of
 $H_N\backslash  SL(4 \vert  N ; {\bf C}) $, but they carry a
representation of $H _N$ which is non-trivial only for a  $U(1)$
subgroup (strictly
$\bf C^*$ in the complex case).
In fact, they transform as \cite {hw2}
\begin{equation}
U (g_1) \phi (X)  = (\rm{sdet}\,[A+ X C])^{- q}
\phi (Xg_1)
\end{equation}
where $q$ is the $U(1)$ weight of $\phi$, that is to say, $\phi$ is a section
of the holomorphic line bundle $L^q$ over anaytic superspace, where $L$ itself
is associated with charge 1. \par
A generic field which depends on the Grassmann even coordinates $y$
and those of
space-time would, when Taylor expanded in $y$, contain an infinite
number of space-time dependent fields.   This is avoided by taking
$\phi$, to be a holomorphic section of $L^q$. Since the internal space is
compact, there are only an finite
number of such sections and so only a finite number of
space-time dependent fields contained in $\phi$.  For example, if
$N = 2$ the internal space is
$U(1)\backslash SU(2)=  CP^1$ and $\phi$ corresponds to a holomorphic section
of $O(q)$,  the $q$th power of the standard line bundle on $CP^1$. It is well
known that there are only
$q+1$ such sections and
$\phi$ may  be expanded around the south pole as
\begin{equation}
\phi = \sum^q_{i = 0} \phi_i (x, \lambda, \pi) y^q
\end{equation}
yielding a finite number of ordinary space-time dependent fields once
we also Taylor expand about $\lambda = \pi = 0.$

Like chiral superfields, analytic superfields have their dilation
weights linearly related to their weights under certain  internal
$U(1)$ transformations.  The $N = 4$ Yang-Mills superfield strength
transforms under  an
induced representation of $SL(4 \vert 4; {\bf C})$ associated with
the  subgroup
$H_4$ and has
$U (1)$ weight one.  We are interested in Green's functions of
observables. We take the observables to be the gauge invariant
quantities $\rm{Tr} W^n$.
The Green's function
\begin{equation}
G = < \rm{Tr} [W (X_1)]^{n_1 } \dots \ \ \rm{Tr} [W (X_j) ]^{n _j} >
\end{equation}
will then obey the superconformal Ward identities.  These imply
that \cite {world} \cite {hw2}
\begin{equation}
G(X_1 \ , \dots, X_p) = \prod _{j=1}^p
(\rm{sdet}\ [A+ X_j C])^{-n_j}
G(X_1^\prime \ , \dots,
X^\prime_p).
\end{equation}
The question we must answer is: does the above equation determine  $G$
up to constants, given that it can only depend analytically on
the $y$'s?

{}From the usual arguments, it is
straightforward to determine the two- and three-point functions using
conformal invariance.  In particular, one finds for the Abelian
theory that \cite {world}
\begin{equation}
G_{1 2} =  < W(X_1) W(X_2)  > \ =  (\rm{sdet}\, X_{1 2})^{- 1} =
{y^2_{1 2} \over x^2_{12}} +  0(\lambda_{1 2} \pi_{1 2})
\end{equation}
where $X_{1 2} = X_1 - X_2$.   Any Green's function can be written
in terms of the product of two-point Green's functions  times a
superconformal invariant, for example
\begin{equation}
< {\rm{Tr}} [W (X_1)]^2 \ {\rm{Tr}} [W (X_2)]^2 \ {\rm{Tr}} [W(X_3)]^3 \
{\rm{Tr}} (W (X_4))^2>
= G^2_{12} G^2_{3 4} \times I\end{equation}
where I is an invariant.

However, unlike the case of chiral superspace,  there are
superconformal invariants in analytic superspace.  It can be shown
\cite {hw2} that they are either of the form
\begin{equation}
{\rm{sdet}\  X_ {ij} \rm{sdet}\  X_{kl} \over \rm{sdet}\
X_{i k}   \rm{sdet} \ X_{j l}}
\end{equation}
or of the form of supertraces of the $X_{ij}'s$ such as
\begin{equation}
\rm{Str} \{X^{-1}_{ij} X_{jk}  X^{-1}_{kl} X_{li}\}
\end{equation}
We refer to these as type I and type II invariants respectively.
The above superconformal invariant
$I$ can then be taken to be  the most general function of  these
invariants subject to the constraint that the Green's function be  an
analytic function of
$y_{i j}.$  Any $N$ point function is an invariant holomorphic section of a
line bundle on $N$
copies of analytic superspace, the particular line bundle being determined by
the $U(1)$ weight of the operators involved. The Green's function will
be determined up to constants if there exist only a finite number of such
sections of the approriate bundle. We are unaware of any general theorems
that classify the number of such sections and have had instead to rely on
an explicit examination of the poles in $y_{ij}$ of the function $I$
to find the restrictions analyticity places upon it.

The  requirement of anaylticity
places very strong constraints on the form of the function $I$.
For example, if  the function $I$ is composed of only type I
invariants,  i.e. super cross-ratios,  then it is easy to convince
oneself that analyticity requires that any Green's function is given
by a sum of terms each of  which is a product of two point functions
multiplied by  arbitary constants. Hence once we make this
restriction the Green's functions are determined up to
constants and take a form reminiscent of free field theory. However, when we
allow for
the possibility of type II invariants the situation becomes
much more complicated.  The calculations one has to do to check analyticity are
long and
complicated, but have been carried out for the  four point Green's
functions of  N=2 analytic operators \cite{hw3}. The result is that such
Green's functions can be completely
determined for operators of charge two and three, but arbitary
functions can occur in the Green's functions of higher charge
operators.
It is likely that  this  calculation can be extended to
higher point Green's functions in
$N=2$  and $N=4$ theories with similar conclusions.
One would also expect to find that the Green's functions
in $N=4$ Yang-Mills theory are determined up to constants by
superconformal invariance alone for a class of sufficiently low
dimension analytic operators.

In some senses it
is to be expected that one cannot determine the Green's functions for
all operators  since this would implicitly require  a definition of
the theory with an action involving operators of  arbitarily high
dimension. However, the Green's functions could be further restricted by
requiring that
they
satisfy physical properties such as crossing and
unitarity. One may also be able to use the bootstrap programme to
determine higher point Green's functions in terms of lower point
Green's functions.

One can also apply the superconformal techniques of this article to
study the operator product expansions of analytic operators in
superconformal field theories
\cite {ope}. In particular, one finds that the $N=4$ energy momentum
tensor supermultiplet has a very simple operator
product expansion which has striking similarities with the
operator product expansion for the energy
momentum tensor in two dimensional spacetime. A heuristic argument for
solvability of $N=4$ Yang-Mills based on the OPE was given in reference
\cite{ope}.

It would be of interest to extend the analysis outlined here to spontaneously
broken superconformal symmetry. If one could calculate some Green's functions
in this phase one might hope to be able to verify the predictions of duality
directly. Finally, it is also possible to study anomalous superconformal Ward
Identities, for example in $N=2$ theories; this was done in \cite{n2} where it
was used to derive the `Matone Identity'\cite{m} for the Seiberg-Witten
prepotential \cite{sw}.

\end{document}